\documentclass[reprint,
 amsmath,amssymb,
 aps,
]{revtex4-2}

\usepackage{graphicx}
\usepackage{dcolumn}
\usepackage{bm}

\usepackage[]{hyperref}  
\usepackage[hyphens]{xurl}
\usepackage[hyphenbreaks]{breakurl}

\begin{document}

\preprint{APS/123-QED}

\title{Comments on ``Final CONUS results on coherent elastic neutrino nucleus scattering at the Brokdorf reactor"}

\author{J.I.\ Collar}
\affiliation{Enrico Fermi Institute, Kavli Institute for Cosmological Physics, and Department of Physics\\
University of Chicago, Chicago, Illinois 60637, USA}

\affiliation{Donostia International Physics Center (DIPC), Paseo Manuel Lardizabal 4, 20018 Donostia-San Sebastian, Spain}

\author{C.M.\ Lewis}
\affiliation{Enrico Fermi Institute, Kavli Institute for Cosmological Physics, and Department of Physics\\
University of Chicago, Chicago, Illinois 60637, USA}

\affiliation{Donostia International Physics Center (DIPC), Paseo Manuel Lardizabal 4, 20018 Donostia-San Sebastian, Spain}



\maketitle


CONUS employs p-type point contact germanium detectors (PPCs, \cite{ppc}) to search for Coherent Elastic Neutrino-Nucleus Scattering (CE$\nu$NS) from reactor (anti)neutrinos. In this context the CE$\nu$NS signal has an endpoint at approximately 0.3 keVee. Microphonic and electronic noise events are strongly dominant for PPCs in this energy regime. Demonstrating a complete rejection of these is crucial for this type of measurement. This has been emphasized in our \cite{ourprl}, which reports on a similar search using a large-mass PPC. Ackermann {\it et al.}\ \cite{acker} claim to exclude the findings in \cite{ourprl}. As described below, the information in \cite{acker} is insufficient to back up this claim. 

Fig.\ 3 in \cite{acker} purports to show the ``measured reactor on and reactor off spectra” for CONUS detector C2, one of three involved. This omits the correction for signal acceptance (SA) near threshold, i.e., does not portray reconstructed spectra that would allow a reader to assess if any background contamination survives in the critical $<0.3$ keVee region of interest (ROI). The CONUS collaboration has nevertheless displayed this missing information in recent presentations. For instance, in slide \#12 at \cite{slides}, where inspection is also facilitated by an expanded energy region. It is apparent there that a rapid rise in the reconstructed spectrum takes place near the 0.21 keVee threshold, with an inflection point at around 0.3 keVee. Applying the SA curve for C2 in \cite{slides} to the data in Fig.\ 3 of Ackermann {\it et al.}\ reproduces this behavior, generating the same reconstructed signal rate of $\sim$100 counts/keV-kg-day at threshold. To our knowledge, no source of radiation is expected to produce this rise \cite{anote}. A recent study of CONUS backgrounds \cite{conusbckg} does not predict such sources. In this situation, it is natural to entertain a concern that a fraction of accepted events below 0.3 keVee might originate in unrejected electronic noise and/or microphonics. As discussed in our work, the latter are highly correlated to PPC cryocooler power. 

Slide \#11 in \cite{slides} shows CONUS detector pulses at 0.3 keVee. These can be used to estimate how similar a 0.21 keVee signal would be to an upwards noise excursion in an energy region where rise-time pulse-shape discrimination (PSD) cuts, described as ``conservative” in \cite{acker}, range from lax to non-existent (see bottom-left panel in this same slide \#11). At this point it is important to emphasize \textbf{ a key issue: unrejected low-energy microphonic events commonly display the same long rise-time of surface events in a PPC.} This is nowhere mentioned by Ackermann {\it et al.}, while acknowledging a large acceptance of long rise-time backgrounds. Microphonic-rejection methods based on time spanned between events, used by CONUS, do not eliminate a necessarily finite 
fraction \cite{0note}  of temporally-isolated signals of this type \cite{ourprl}. Hence the importance of a truly conservative rise-time cut \cite{ncc1701} and the reality of the concern expressed above.

Ackermann {\it et al.}\ assert ``highly stable conditions” during these runs. However, inspection of their Fig.\ 2 reveals a noticeable increase in both detector electronic noise and cryocooler power during reactor OFF periods, compared to reactor ON runs. A recently added appendix to \cite{acker} confirms that this extends to all detectors. While these variations might seem modest at first sight, the CONUS collaboration has quantified the magnitude of microphonic contamination in their detectors in Figs.\ 3.6a and 3.21 of \cite{thesis}: at 0.21 keVee, the raw count rate prior to any cuts against microphonics is of order 6,000-40,000 counts/keV-kg-day, or up to several thousand times the irreducible background shown in Fig.\ 3 of Ackermann {\it et al.}\ The ``almost linear correlation" between microphonic event rate below 0.28 keVee and minuscule changes in cryocooler power, smaller than those in \cite{acker}, can be ascertained in Fig.\ 3.22 of \cite{thesis}. In these conditions, \textbf{even a small remanent noise contamination would be able to increase the reactor OFF event rate in the $<0.3$ keVee region with respect to reactor ON periods, encumbering the observation of a CE$\nu$NS excess in the ON-OFF residual or leading to tighter constraints on it.} The rise in reconstructed spectra discussed above indicates that this contamination can be sizeable, as one third or more of the event rate at the intended 0.21 keVee analysis threshold originates in long rise-time backgrounds. 

Ackermann {\it et al.}\ offer in their Fig.\ 2 two parameters supposed to exemplify the stability of their detectors. The first is the reset rate of the transistor-reset preamplifier (TRP). This is a mere measure of the electronic leakage current of a PPC, which is typically highly constant over periods of up to a few years in conditions of relatively stable ambient temperature \cite{ourprl}. This parameter provides  no information on the leading source of microphonic noise contamination, which are unrelated mechanical vibrations in the cryocooler, correlated to its power consumption and in turn even to minor changes in ambient temperature (this last is mentioned and noticeable in the information provided in \cite{acker}). The second parameter offered as an alleged measure of stability is the energy resolution of a 10.4 keV electron capture decay peak. As discussed in \cite{eres}, at this energy the resolution of a PPC is mainly defined by the intrinsically-stable dispersion in the statistics of information carriers (electron-hole pairs), which derives exclusively from the amount of energy deposited by the decay. Needless to say, this is a time-independent constant. Conversely, detector noise becomes entirely dominant in defining the resolution at the lower energy of the CE$\nu$NS ROI. In conclusion, these two parameters may communicate a superficial impression of PPC stability, but in reality no proof of signal rate stability in the CE$\nu$NS ROI and of its lack of correlation to dominant PPC noise sources is presently offered by Ackermann {\it et al.}\  \cite{bnote}. This is in contrast to Fig.\ 2 of our \cite{ourprl}, where the evolution of the accepted irreducible event rate in the relevant 0.2-0.3 keVee ROI and that of the rejected backgrounds in the same region was provided. Their correlation to noise sources was examined, excluded for the first and established for the second. These pertinent parameters and the mode of correlation analysis that we employed are also available to Ackermann {\it et al.}\ In view of the strong concerns expressed here, they must be provided to support their claims. To quote from \cite{ourprl}, ``This type of correlation analysis is of crucial importance prior to a CE$\nu$NS search using PPCs, in view of the dominance of microphonic noise in the ROI [27]” (where ref.\ 27 is the CONUS thesis pointed out in these Comments \cite{thesis}). 

Even if these deficiencies and omissions can be corrected to substantiate the central tenet in Ackerman {\it et al.}, multiple other issues remain: {\it i)} it is claimed that in our \cite{ourqf} ``the Migdal effect was proposed as a physical explanation for the enhancement” in a low-energy quenching factor (QF) for nuclear recoils in germanium. This description is incomplete: this effect was examined as just one of several possible ``new physical process (or processes)”. Those were listed as ``alternative paths leading to an enhanced ionization yield” and cited as references 71-73 in \cite{ourqf}. {\it ii)} A theoretical study (ref.\ 25 in \cite{acker}) is mentioned as ``demonstrating” that the Migdal effect is negligible in the present context. Said study fails to recognize the relevance of ultra-low energy secondary recoils emphasized in Sec.\ VII of \cite{ourqf}. Other studies argue for an enhanced importance of Migdal for such recoils in semiconductors (e.g., ref. 40 in \cite{ourqf}, or more recently \cite{migdal}). This area of knowledge is still in a theoretical and experimental state of flux, one that is na{\"i}vely described by Ackermann {\it et al.}\ {\it iii)} Mention is made that ``good agreement between experimental data and the Lindhard prediction was found recently”, citing \cite{theirqf} (criticized in \cite{critique}) and unpublished thesis work. For completeness, Ackerman {\it et al.}\ should comment on most recent experimental work \cite{kavner} that points to the contrary and supports the QF results in \cite{ourqf}. {\it iv)} Finally, the YBe QF cubic fit showed in Fig.\ 4 of Ackermann {\it et al.}, employed to extract their CE$\nu$NS rate prediction in Table II departs from the QF data that we used and provided in \cite{ourprl}. No justification is given for this \cite{cnote}. Rising faster at low recoil energy, this manipulated QF  can cause a somewhat inflated rate prediction \cite{dnote}, facilitating the claim of (marginal) exclusion of our findings. It should be noted that the best-fit CE$\nu$NS signal found in \cite{ourprl} lays in close vicinity to the predictions from this YBe QF model (Fig.\ 4 in \cite{ourprl}). In this last respect, both this YBe QF model and that proposed by CONUS in \cite{theirqf} (Lindhard $\kappa\!=\!0.162$) generate statistically compatible predictions in Table II of \cite{acker}, even when using this inaccurate cubic fit. There is a manifest dissonance between this fact and the conclusions in Ackermann {\it et al.}, one that seems difficult to defend.


\bibliography{apssamp}

\begin{thebibliography}{18}
\bibitem [{ano()}]{anote}%
  \BibitemOpen
  \href@noop {} {}\bibinfo {note} {M-shell electron capture in $^{71}$Ge is negligible for the present CONUS neutron background.}\BibitemShut {Stop}%
\end{thebibliography}

\end{document}